\documentclass[12pt,a4paper]{article}
\usepackage{amsfonts,latexsym}
\usepackage{amsmath,amssymb}
\usepackage{graphicx,color}

\numberwithin{equation}{section}

\renewcommand{\thefootnote}{\fnsymbol{footnote}}
\newcommand{\nn}{\nonumber}
\begin{document}

\vspace{12mm}

\begin{center}
{{{\Large {\bf Instability of rotating black hole \\ in a limited form of  $f(R)$ gravity }}}}\\[10mm]

{Yun Soo Myung\footnote{e-mail address: ysmyung@inje.ac.kr}} \\[10mm]
{ Institute of Basic Sciences and School of Computer Aided Science, Inje University,  Gimhae 621-749, Korea\\[0pt]}

\end{center}
\vspace{2mm}

\begin{abstract}
We  investigate the stability of $f(R)$-rotating (Kerr) black hole
obtained from a limited form of  $f(R)$ gravity. In order to avoid
the difficulty of handling fourth order linearized equations, we
transform this form of $f(R)$ gravity into the scalar-tensor theory
by introducing two auxiliary scalars.
 In this case, the linearized curvature scalar equation leads to a massive scalaron equation.
   It turns out that the  $f(R)$-rotating black hole is unstable  because of
   the superradiant instability known as the black-hole bomb idea.

\end{abstract}
\vspace{5mm}

{\footnotesize ~~~~PACS numbers: }

%{\footnotesize ~~~~Keywords: Classical Theories of Gravity, Spacetime Singularities, Black Holes in String Theory}

\vspace{1.5cm}

\hspace{11.5cm}{Typeset Using \LaTeX}
\newpage
\renewcommand{\thefootnote}{\arabic{footnote}}
\setcounter{footnote}{0}

%%%% Introduction %%%%

\section{Introduction}
Modified gravity theories, $f(R)$ gravities~\cite{NO,sf,NOuh} have
much attentions as one of strong candidates for explaining the
current accelerating universe~\cite{SN}.  $f(R)$ gravities can be
considered as Einstein gravity  with an additional scalar field. For
example,  the metric-$f(R)$ gravity is equivalent to the
$\omega_{\rm BD}=0$ Brans-Dicke (BD) theory with the
potential~\cite{FT}.  It was shown that the equivalence principle
test allows  $f(R)$ gravity models that are nearly indistinguishable
from the ${\rm \Lambda}$CDM model in the background universe
evolution~\cite{PS}.  However, this does not imply that there is no
difference in the dynamics of perturbations~\cite{CENOZ}.  It is
worth noting  that the perturbation distinguishes between Einstein
and $f(R)$ gravities.

In order for $f(R)$ gravities to be acceptable, they obey certain
minimal requirements for theoretical viability~\cite{sf,FT}. Three
important requirements are included: (i) they  possess the correct
cosmological dynamics, (ii) they are free from instabilities,
tachyon and ghosts~\cite{CLF,PS}, (iii) they attain the correct
Newtonian and post-Newtonian limits.

On the other hand, the Schwarzschild-de Sitter black hole was
obtained for a constant curvature scalar from $f(R)$
gravity~\cite{CENOZ}. A black hole solution was obtained from $f(R)$
gravities by requiring the negative constant curvature scalar
$R=R_0$~\cite{CDM}. For $1+f'(R_0)>0$, this black hole is similar to
the Schwarzschild-AdS (SAdS) black hole. In order to obtain the
(constant curvature) black hole solution from $f(R)$ gravity coupled
to the matters of the Maxwell~\cite{CDM} and Yang-Mills
fields~\cite{MMS}, the trace of its stress-energy tensor
$T_{\mu\nu}$ should be zero. Interestingly, it was pointed out that
the Kerr solution could be obtained from a limited form of  $f(R)$
gravity~\cite{kerr}. Later on, it was argued  that perturbed Kerr
black hole can discriminate Einstein and $f(R)$ gravities~\cite{BS}.

 All astrophysical black holes
belong to the rotating black hole. A black hole solution should be
stable against the external perturbations because it stands as a
physically realistic object~\cite{KZ}.  Studies of stability of Kerr
black hole are not as straightforward, because it is axially
symmetric black hole and thus, the decoupling process seems to be
complicated. The Kerr black hole has been proven to be stable
against gravitational fields~\cite{PT,TP,Whit} and  massless
scalar~\cite{DI}. However, there exist unstable modes when
considering a massive scalar due to the superradiance~\cite{ZE}.
Furthermore, it seems that the stability analysis of $f(R)$-rotating
black hole is a formidable task because $f(R)$ gravity contains
fourth  order derivatives in the linearized equations~\cite{BS}.

In this work, we investigate the stability of $f(R)$-rotating (Kerr)
black hole arisen from a limited form  (\ref{fform}) of  $f(R)$
gravity~\cite{kerr}. This work will be interesting because if $f(R)$
gravity is considered as a viable theory, it is responsible to
explain the present accelerating universe as well as the formation
of rotating black holes.  We transform the  form  of $f(R)$ gravity
into the scalar-tensor theory to avoid fourth order derivative terms
by introducing two auxiliary scalars. Then, the linearized curvature
scalar equation becomes a massive scalaron equation, indicating that
all linearized equations are second order. Using the stability
analysis of Kerr black hole in the massive Klein-Gordon equation, we
show clearly that the $f(R)$-rotating black hole is unstable against
the scalaron perturbation.

\section{Perturbation of $f(R)$ black holes}

We start with $f(R)$ gravity without any matter fields whose action
is given by
\begin{eqnarray}
S_{f}=\frac{1}{2\kappa^2}\int d^4 x\sqrt{-g} f(R),\label{Action}
\end{eqnarray}
where $\kappa^2=8\pi G$.  The Einstein equation  takes the form
\begin{eqnarray} \label{equa1}
R_{\mu\nu} f'(R)-\frac{1}{2}g_{\mu\nu}f(R)+
\Big(g_{\mu\nu}\nabla^2-\nabla_{\mu}\nabla_{\nu}\Big)f'(R)=0,
\end{eqnarray}
where ${}^{\prime}$ denotes the differentiation with respect to its
argument.  It is well-known that (\ref{equa1}) has a solution with
constant curvature scalar $R=\bar{R}$. In this case,  (\ref{equa1})
can be written as
\begin{eqnarray} \label{equ1}
\bar{R}_{\mu\nu} f'(\bar{R})-\frac{1}{2}g_{\mu\nu}f(\bar{R})=0,
\end{eqnarray}
and thus,  the trace of (\ref{equ1}) becomes
\begin{eqnarray}
\bar{R}=\frac{2f(\bar{R})}{f'(\bar{R})}\equiv 4\Lambda_f
\label{eqCR}
\end{eqnarray}
with $\Lambda_f$ the cosmological constant due to the $f(R)$
gravity.  Substituting this expression into (\ref{equ1}), one
obtains the Ricci tensor
\begin{equation}
\bar{R}_{\mu\nu}=\frac{f(\bar{R})}{2f'(\bar{R})}\bar{g}_{\mu\nu}=\Lambda_f
\bar{g}_{\mu\nu}.
\end{equation}
In order to find the Kerr black hole solution, we have to choose a
non-pathologically functional form of $f(R)$ as~\cite{kerr}
\begin{equation}
\label{fform} f(R)=a_1R+a_2R^2+a_3R^3+\cdots
\end{equation}
which is surely a Talyor series around $R=\bar{R}=0$. We should
mention that this form of $f(R)$ gravity is not general,  but a
small subset of $f(R)$ gravities.   We check that
$f(0)=0,~f'(0)=a_1,~f''(0)=2a_2$ at $R=\bar{R}=0$, which provides
either the Schwarzschild or  Kerr black hole solution when choosing
$\Lambda_f=0$.

 In this work, we use the Boyer-Lindquist coordinates
to represent an axis-symmetric Kerr black hole solution with mass
$M$ and angular momentum $J$~\cite{Kerrsol},
\begin{eqnarray}
ds^2_{\rm Kerr} \!\!&=&\!\! -\left ( 1-\frac{2Mr}{\rho^2}\right )dt^2
-\frac{2Mr a\sin^2\theta}{\rho^2}\, 2 dt d\phi +\frac{\rho^2}{\Delta}\,dr^2 \nonumber \\
& &  +\rho^2 d\theta^2+\left ( r^2+a^2+\frac{2Mr
a^2\sin^2\theta}{\rho^2}\right )\sin^2\theta\, d\phi^2 \, \nonumber \\
& &
 \label{Kerr}
\end{eqnarray}
with
\begin{eqnarray}
\Delta=r^2+a^2-2Mr\,, \qquad  \rho^2=r^2+a^2 \cos^2\theta \,,\qquad a=\frac{J}{M}.
 \label{metric parameters}
\end{eqnarray}
In the limit of $a\to 0$, (\ref{Kerr}) recovers the Schwarzschild
black hole, while $a\to1$ goes to the extremal Kerr black hole. The
zeros of $\Delta$, two horizons are located at
\begin{equation}
r_\pm=M\pm \sqrt{M^2-a^2}
\end{equation}
and the angular velocity at the event horizon takes the form
\begin{equation}
\Omega=\frac{a}{2 M r_+}. \label{hav}
\end{equation}

Now we introduce the perturbation around the Kerr
black hole to study stability of the black hole
\begin{eqnarray} \label{m-p}
g_{\mu\nu}=\bar{g}_{\mu\nu}+h_{\mu\nu}.
\end{eqnarray}
Hereafter we denote the background quantities with the ``overbar''.
Then, the linearized equation around $f(R)$-rotating black hole
becomes
\begin{eqnarray}
&&\bar{\nabla}^{\rho}\bar{\nabla}_{\mu}h_{\nu\rho}+
\bar{\nabla}^{\rho}\bar{\nabla}_{\nu}h_{\mu\rho}-\bar{\nabla}^2h_{\mu\nu}
-\bar{\nabla}_{\mu}\bar{\nabla}_{\nu}h
-\bar{g}_{\mu\nu}\Big(\bar{\nabla}^{\alpha}\bar{\nabla}^{\beta}h_{\alpha\beta}
-\bar{\nabla}^2h\Big)\nn\\ &&\hspace*{8em}
+\Big[\frac{2}{3m^2_f}\Big]\Big(\bar{g}_{\mu\nu}\bar{\nabla}^2-\bar{\nabla}_{\mu}
\bar{\nabla}_{\nu}\Big) \Big(
\bar{\nabla}^{\alpha}\bar{\nabla}^{\beta}h_{\alpha\beta}
-\bar{\nabla}^2h\Big)=0\label{leq1}
\end{eqnarray}
with the mass squared $m^2_f$ defined by
\begin{equation}
m^2_f=\frac{f'(0)}{3f''(0)}.
\end{equation}
Taking the trace of $(\ref{leq1})$ with $\bar{g}^{\mu\nu}$,  one
has the fourth order equation for $h_{\mu\nu}$
\begin{eqnarray}
\Big(\bar{\nabla}^2-m^2_f\Big)\delta R(h)=0 \to
\Big(\bar{\nabla}^2-m^2_f\Big)\Big(
\bar{\nabla}^{\alpha}\bar{\nabla}^{\beta}h_{\alpha\beta}
-\bar{\nabla}^2h\Big)=0. \label{leq2}
\end{eqnarray}
At this stage, we note that it is not easy to make a further
progress on  the perturbation analysis  because there exist fourth
order derivatives.  We mention  that  for the Einstein gravity with
$f(R)=R$, $f'(0)=1$ and $f''(0)=0$. In this case, one finds  the
equation for linearized curvature scalar: $\delta R(h)=0$, which
means that $\delta R(h)$  is not a physically propagating mode.
Actually, this equation leads to one constraint as
\begin{equation} \label{cons}
\bar{\nabla}^{\alpha}\bar{\nabla}^{\beta}h_{\alpha\beta}
=\bar{\nabla}^2h
\end{equation}
which will also be recovered from the transverse gauge.

 Up to now,
we did not fix any gauge. We would like to comment on the linearized
equation by  choosing the Lorentz gauge~\cite{FH}
\begin{equation}
\bar{\nabla}_{\nu}h^{\mu\nu}=\frac{1}{2}\bar{\nabla}^{\mu}h.
\end{equation}
Under this gauge-fixing, the linearized equation (\ref{leq1}) takes
a simple  form
\begin{eqnarray} \label{bs}
\bar{\nabla}^2\tilde{h}_{\mu\nu}+2\bar{R}_{\mu\rho\nu\sigma}\tilde{h}^{\rho\sigma}
+\frac{1}{3m^2_f}\Big(\bar{g}_{\mu\nu}\bar{\nabla}^2-\bar{\nabla}_{\mu}
\bar{\nabla}_{\nu}\Big)\bar{\nabla}^2\tilde{h}=0
\end{eqnarray}
with the trace-reversed perturbation
$\tilde{h}_{\mu\nu}=h_{\mu\nu}-h\bar{g}_{\mu\nu}/2$ and
$1/3m^2_f=\lambda$~\cite{BS}. This equation was importantly used to
mention that perturbed Kerr black holes obtained from  $f(R)$
gravity can probe deviations from the Einstein gravity~\cite{BS}.
Even though equation (\ref{bs}) is simpler than (\ref{leq1}), it is
a non-trivial task to decouple odd and even perturbations around the
Kerr black hole, arriving at two fourth order equations hopefully.
Crucially, we  do not know how to solve the fourth order equation
(\ref{bs}) around the $f(R)$-rotating black hole. On the other hand,
we may choose the transverse gauge which works well for studying the
graviton propagations on the the Minkowski and AdS$_4$ spacetime
backgrounds~\cite{GT,Myungf}
\begin{equation}
\bar{\nabla}_{\mu}h^{\mu\nu}=\bar{\nabla}^{\nu}h,
\end{equation}
which leads to (\ref{cons})
 when operating $\bar{\nabla}$ on both sides. Using the
 relation (\ref{cons}), one immediately finds that the effect of $f(R)$ gravity
 [$2/3m^2_f$-term in (\ref{leq1})] disappears because of $\delta R(h)=0$, leading to the Einstein gravity.
  Hence, the gauge-fixing for stability analysis of the $f(R)$-rotating black hole
  differs  from that  for the graviton propagations on
 the AdS, dS, and Minkowski spacetimes~\cite{FH}.
 We note that choosing an appropriate  gauge-fixing cannot resolves the difficulty with $f(R)$ gravities. It seems that the best way
 to resolve the difficulty confronting with the fourth order
 equation is to translate the fourth order equation into the second
 order equations by introducing auxiliary scalar fields. In other words,
 we must  make a transformation from the limited form of $f(R)$ gravity to the
 scalar-tensor theory (like Brans-Dicke theory) to perform  the stability analysis of $f(R)$-rotating black hole.

\section{Perturbation of  the scalar-tensor theory}

In this section, we will develop the perturbation analysis around
the $f(R)$-rotating black holes (\ref{Kerr}) in the different frame,
the scalar-tensor theory. Introducing two auxiliary fields $\phi$
and $A$,  one can rewrite the  action (\ref{Action})
as~\cite{Olmo,Olmof}
\begin{eqnarray}
S_{st}=\frac{1}{2\kappa^2}\int d^4
x\sqrt{-g}\Big[\phi\left(R-A\right)+f_A(A)\Big]. \label{ActionfA}
\end{eqnarray}
Varying for the fields $\phi$ and $A$ leads to two equations
\begin{eqnarray}
R=A,~~\phi=f'_A(A).\label{eomA}
\end{eqnarray}
 Note that using (\ref{eomA}),  the  action (\ref{ActionfA})
recovers  the original action (\ref{Action}). On the other hand, the
equation of motion for the metric tensor  can be obtained by
\begin{eqnarray} \label{equa}
f'_A(A)
R_{\mu\nu}-\frac{f_A(A)}{2}g_{\mu\nu}
+\Big(g_{\mu\nu}\nabla^2-\nabla_{\mu}\nabla_{\nu}\Big)f'_A(A)=0.\label{eomg}
\end{eqnarray}
Considering a constant curvature scalar $R=\bar{R}=\bar{A}$ together with
$\bar{\phi}=f'_A(\bar{A})={\rm const}$, Eq.(\ref{eomg}) becomes
\begin{equation}
f'_A(\bar{A})\bar{R}_{\mu\nu}-\frac{1}{2}\bar{g}_{\mu\nu}f_A(\bar{A})=0.\label{eomg1}
\end{equation}
Taking the trace of (\ref{eomg1}) leads
to
\begin{eqnarray}
\bar{R}=\frac{2f_A(\bar{A})}{f'_A(\bar{A})}\equiv 4\Lambda_A.
\label{eqCR}
\end{eqnarray}
Substituting this expression into (\ref{eomg1}), one finds the
Ricci tensor which determines the maximally symmetric Einstein spaces including Minkowski space
\begin{equation}
\bar{R}_{\mu\nu}=\frac{1}{2}\frac{f_A(\bar{A})}{f'_A(\bar{A})}\bar{g}_{\mu\nu}=\Lambda_A
\bar{g}_{\mu\nu}.
\end{equation}
At this stage, we introduce a specific form  of $f_A(A)$ inspired from
(\ref{fform})~\cite{kerr}
\begin{eqnarray}
\label{f-form}f_A(A)=a_1 A+a_2 A^2+a_3 A^3\cdots,
\end{eqnarray}
where $a_1,a_2,a_3,\cdots$ are arbitrary constants. Their mass
dimensions are $[a_1]=0,~[a_2]=-2$ because $[f_A(A)]=2,~[A]=2,$ and
$[\phi]=2$. In this work,  we confine ourselves to the
asymptotically flat spacetimes  with $\Lambda_A=0$ which
accommodates the $f(R)$-rotating black hole (\ref{Kerr}). In this
case, we have to choose $\bar{A}=0$ and thus,
\begin{equation}
f_A(0)=0,~~ f'_A(0)=a_1>0,~f''_A(0)=2a_2>0 \label{mink}.
\end{equation}
Now we are in a position to  study the perturbation around the
Kerr black hole (\ref{Kerr}).  In addition to (\ref{m-p}), from (\ref{eomA}), we have
\begin{equation}
\bar{R}+\delta R(h)=\bar{A}+\delta A,~~\bar{\phi}+\delta \phi=f'_A(\bar{A})+f''_A(\bar{A})\delta A,
\end{equation}
which leads to
\begin{equation} \label{pr-p}
\delta R(h)\to \delta A,~~\delta \phi \to f''_A(\bar{A})\delta A.
\end{equation}
Thus,  instead of $\delta R(h)$-$\delta \phi$, we use  $\delta A$ as
a perturbed field  in addition to $h_{\mu\nu}$. We mention that if
one uses $\delta R(h)$-$\delta \phi$, one gets the same linearized
equations. For this,  see the appendix of Ref.\cite{MMSf}.

Using (\ref{pr-p}), the linearized equations to (\ref{equa}) reduces
to
\begin{eqnarray}
\delta
R_{\mu\nu}(h)=\Big[\frac{1}{3m^2_A}\Big]\bar{\nabla}_{\mu}\bar{\nabla}_{\nu}\delta
A+\frac{1}{6}\bar{g}_{\mu\nu}\delta A\label{eqRmunu1},
\end{eqnarray}
where the scalaron mass squared is given by
\begin{equation}
m^2_A=\frac{f'_A(0)}{3f''_A(0)}.
\end{equation}
Taking the trace of (\ref{eqRmunu1}) and using (\ref{pr-p}) leads to the massive scalaron equation
\begin{equation}
\Big(\bar{\nabla}^2-m_A^2\Big)\delta A=0.
 \end{equation}
Since the mass dimension of  the linearized scalaron is two
($[\delta A]=2$), it would be better to write the canonically
linearized equations by introducing a dimensionless scalaron $\delta
\tilde{A}$ defined by
\begin{equation}
\delta \tilde{A}=\frac{\delta A}{3m^2_A}. \end{equation} Finally, we
arrive at two linearized equations
\begin{eqnarray}
&&\Big(\bar{\nabla}^2-m_A^2\Big)\delta \tilde{A}=0,\label{gtrf} \\
&& \delta R_{\mu\nu}(h)=\bar{\nabla}_{\mu}\bar{\nabla}_{\nu}\delta
\tilde{A}+\Big[\frac{m^2_A}{2}\Big]\bar{g}_{\mu\nu}\delta \tilde{A},
\label{eqRmunu1f} \end{eqnarray} which are our main equations to
carry out the stability analysis of $f(R)$-rotating black hole. It
is a nontrivial task to perform the stability analysis of
$f(R)$-rotating black hole  based on the metric-perturbing equation
(\ref{eqRmunu1f}) since the decoupling process seems to be
complicated. However, we expect that the right hand side of scalaron
coupling terms in (\ref{eqRmunu1f})  will not change the stability
of $f(R)$-rotating black hole~\cite{MMSf,MMSfa}. In this work, we
focus on the scalaron equation (\ref{gtrf}) because it shows the
feature of $f(R)$ gravities.

\section{Stability analysis of $f(R)$-rotating black hole}

Considering the axis-symmetric background (\ref{Kerr}), it is
convenient to separate the scalaron into~\cite{Teu}
\begin{equation}
\delta \tilde{A}(t,r,\theta,\phi)=e^{-i\omega t + i m \phi} S^m _l(\theta)
R(r)\,, \label{separation}
\end{equation}
where $S^m _l(\theta)$ are spheroidal angular functions, and the
azimuthal number $m$ takes on  (positive or negative) integer
values. Also, it is enough to consider positive $\omega$'s in
(\ref{separation}). Plugging this into the scalaron (Klein-Gordon)
equation (\ref{gtrf}), we get the  angular and radial wave equations
for $S^m _l(\theta)$ and $R(r)$ as
\begin{eqnarray}
& & \frac{1}{\sin \theta}\partial_{\theta}\left ( \sin \theta
\partial_{\theta} S^m _l \right )  \nonumber \\
& &\hspace{1cm}+
\left [  a^2 (\omega^2-m^2_A) \cos^2
\theta-\frac{m^2}{\sin ^2{\theta}}+A_{lm} \right ]S^m _l =0\,,
\label{wave eq separated1}
\\
& & \Delta\partial_r \left ( \Delta \partial_r R \right )+ {\bigl [}
\omega^2(r^2+a^2)^2-4M a m
\omega r +a^2 m^2 \nonumber \\
& & \hspace{3.2cm} - \Delta (a^2\omega^2+ m^2_Ar^2+A_{lm}) {\bigr ]} R=0\,,
 \label{wave eq separated}
\end{eqnarray}
where $A_{lm}$ is the separation constant that allows the split of
the wave equation~\cite{BPT,HH}
\begin{eqnarray}
A_{lm}=l(l+1)+\sum^\infty_{k=1}c_ka^{2k}(m^2_A-\omega^2)^k\,.
 \label{eigenvalues}
\end{eqnarray}
The radial Teukolsky equation takes the Schr\"odinger form
\begin{equation}
-\frac{d^2\tilde{R}}{dr^{*2}}+V(r,\omega)\tilde{R}=\omega^2{\tilde{R}},~~\tilde{R}=\sqrt{r^2+a^2}R,
\end{equation}
where the tortoise $r_*$ coordinate is defined by $dr^*=
\frac{r^2+a^2}{\Delta}dr$ and the $\omega$-dependent potential
$V(r,\omega)$ is given by
\begin{eqnarray}
V(r,\omega)&=&\frac{\Delta
m^2_A}{r^2+a^2}+\frac{4Mram\omega-a^2m^2+\Delta[A_{lm}+(\omega^2-m^2_A)a^2]}{(r^2+a^2)^2}
 \nonumber \\
 &+&\frac{\Delta(3r^2-4Mr+a^2)}{(r^2+a^2)^3}-\frac{3\Delta^2r^2}{(r^2+a^2)^4}.
\end{eqnarray}
Its asymptotic forms are given by
\begin{equation}
V \to \omega^2-m^2_A,~~r^*\to \infty(r\to\infty),
\end{equation}
\begin{equation}
V \to (\omega-m\Omega)^2,~~r^*\to -\infty(r\to r_+).
\end{equation}
 Near  the horizon at
$r=r_+$ and spatial infinity at $r=\infty$, the scalaron takes the
form
\begin{eqnarray}
\tilde{R} &=& {\cal T} e^{- i(\omega-m\Omega)
r^*}\,\,,\,\,r^* \to -\infty \label{bc2}\\
\tilde{R} &=& e^{-i\sqrt{\omega^2-m^2_A}r^*}+{\cal
R}e^{i\sqrt{\omega^2-m^2_A}r^*},~~r^*\rightarrow \infty \label{bc1}
\end{eqnarray}
with the ${\cal T}({\cal R})$ the  transmission (reflection)
amplitudes. Here we must assume that $\omega$, for the bound case of
$\omega<m_A$, lies on the physical sheet  of $0\le {\rm arg}
(\omega^2-m^2_A)^{1/2} <\pi$.
 Requiring ingoing waves upon  a rotating
black hole whose angular velocity $\Omega$ is given by (\ref{hav}),
one must impose an ingoing (negative) group velocity $ v_{\rm gr}$
for the wave packet. Then, we choose  an ingoing mode near the even
horizon as
\begin{equation}
[e^{-i\omega t}\tilde{R}]_{\rm in}= {\cal T} e^{-i\omega t}e^{-
i(\omega-m\Omega) r_*}.\end{equation} Since $V(r,\omega)$ is real,
the Wrongskian $W(\tilde{R},\tilde{R}^*)$ of the complex conjugate
solutions of $\tilde{R}$ and $\tilde{R}^*$ satisfies~\cite{KZ}
\begin{equation}
i\frac{d}{dr^*} W(\tilde{R},\tilde{R}^*)=0
\end{equation}
which implies that
\begin{equation}
|{\cal R}|^2=1+\Bigg[\frac{m\Omega}{\omega}-1\Bigg]|{\cal T}|^2.
\end{equation}
Here, if the frequency $\omega$ of the incident wave satisfies the
inequality
\begin{equation}
\omega < m \Omega \,, \label{super}
\end{equation}
one has $|{\cal R}|>1$, implying that the reflected (scattered) wave
is being amplified~\cite{Teu,TP}. Thus, in this superradiance
regime, waves appear as outgoing to an observer at infinity, and
energy radiated away to infinity actually exceed the energy present
in the initial perturbation. Feeding back the amplified scattered
waves, one may gradually extract  the rotational energy of the
$f(R)$-rotating black hole. Press and Teukolsky have suggested to
use this mechanism to the black-hole bomb~\cite{PT}. If one
surrounds the black hole by a reflecting mirror, the wave will
bounce back and forth between black hole and mirror, amplifying
itself each time. To this end, nature may provide its own
mirror~\cite{BPT,CDLY}.

It is well known that if one considers a massive scalar field with
mass $\mu$ scattered off a rotating black hole, then for $\omega <
\mu$, the superradiance has unstable modes because  the mass term
effectively works as a reflecting
mirror~\cite{DDR,ZE,Det,FN,SK,Dol}. As was shown in Fig. 15 of
Ref.\cite{KZ}, a shape (ergoregion-$\sim$-mirror) of potential
$V(r,\omega)$ has the local maximum as well as the local minimum far
from the black hole which generates a secondary reflection of the
wave the reflected from the potential barrier. This implies that the
meta-stable bound states of $Mm_A\sim r_+/\lambda_c$ with
$\lambda_c$ the Compton wavelength of the massive particle could be
formed in the valley of the local minimum.  The secondary reflected
wave  will be reflected again at the far region. Since each
scattering off the barrier in the superradiant region increases the
amplitude of the wave, the process of reflections will continue with
the increased energies of waves, leading to an instability. In our
case, the mass $m_A$  of scalaron works as a reflecting mirror.
Considering a massive scalaron with mass $m_A$ scattered off
$f(R)$-rotating black hole, then for $\omega < m_A$, the mass term
effectively works as a reflecting mirror.   Similarly, the Kerr
black hole was shown to be unstable when choosing
$f(R)=R+hR^2$~\cite{HO} which is included as a limited form
(\ref{fform}) when setting all $a_n=0$ for $n\ge 3$, $a_1=1$, and
$a_2=h$.

 The instability time scale $\tau$  associated with the dynamics of a massive scalar
  was restricted to two limiting cases:
  $Mm_A \ll 1$ provides  $\tau \simeq 24(a/M)^{-1}(Mm_A)^{-9}(GM/c^3)$~\cite{ZE},
   while $Mm_A \gg 1$ indicates $\tau \sim 10^{7}e^{1.84 Mm_A}(GM/c^3)$~\cite{Det}.
    Recently, the authors~\cite{Dol} have shown that for  $Mm_A\le 0.42$ and $a=0.99$,
    the maximal instability growth rate    is given by $\tau^{-1} \simeq 1.5 \times
    10^{-7}(GM/c^3)^{-1}$.

\section{Discussions}

It was suggested that perturbed Kerr black holes discriminates
Einstein and $f(R)$ gravities~\cite{BS}. To this end, we have made a
significant progress if the scalaron  approach (scalar-tensor
theory) represents truly the
 $f(R)$ gravity including the fourth order differential
equation.

It turned out that the $f(R)$ black hole  is stable against external
perturbations if the scalaron does not have a tachyonic mass
($m^2_A=f'_A(0)/3f''_A(0)> 0,~f_A''(0)>0$)~\cite{MMSf}. This is
consistent with other perturbation analysis: the Dolgov-Kawasaki
instability with $f''(R)<0$ in cosmological perturbations~\cite{DK}.
For the $f(R)$-AdS black hole, its stability is guaranteed  if the
scalaron mass squared $\tilde{m}^2_A$ satisfies the
Breitenlohner-Freedman bound ($\tilde{m}^2_A\ge m^2_{\rm
BF}=-9/4\ell^2$)~\cite{MMSfa}.
 However,
for $f(R)$-rotating  black hole, it is  unstable  even for the
positive  scalaron mass squared  $m^2_A>0$ because of its
superradiant instability (black hole bomb). We would like to stress
that this is a meaningful  result for the rotating black hole
obtained from a limited form (\ref{fform}) of $f(R)$ gravity.

Given the results for the e-folding time
$\tau=10^7e^{1.84Mm_A}(GM/c^3)$ for $Mm_A\gg1$, when might the
instability be significant~\cite{ZE,Dol}? One assumes  that if  the
scalaron is a pion around a solar-mass black hole, $Mm_A\sim
10^{18}$ with $M \simeq 10^{30}$kg.  Then, one concludes that the
instability growth rate $\tau^{-1}$  is not significant for
astrophysical black holes, unless there exists an unknown particles
with a tiny but nonzero rest mass. That is, the instability growth
rate $\tau^{-1}$ is always small for the standard model particles,
in compared to the decay rate of particle or  the Hawking
evaporation rate of the black hole. However, the instability may be
important for a pion around a small primordial black hole ($M\le
10^{12}$kg).

Finally,  $f(R)$ theories are supposed to be viable only when the
matter is present outside the star/black hole so that the Chameleon
mechanism could take place. However, in order to set matter (even a
small amount of matter) outside the star/black hole, it will change
the results of analysis for the $f(R)$ theories, as the scalaron
$\delta A$ becomes massive.

In conclusion, we have shown that  the $f(R)$-rotating black hole
arisen from a limited form (\ref{fform}) of $f(R)$ gravity is
unstable because of
   the superradiant instability known as the black-hole bomb idea.

 \vspace{1cm}
{\bf Acknowledgments}

This work was supported by the National Research Foundation of Korea
(NRF) grant funded by the Korea government (MEST) (No.2010-0028080).

\end{document}